\documentclass[fleqn,usenatbib]{mnras}

\usepackage{newtxtext,newtxmath}

\usepackage[T1]{fontenc}

\DeclareRobustCommand{\VAN}[3]{#2}
\let\VANthebibliography\thebibliography
\def\thebibliography{\DeclareRobustCommand{\VAN}[3]{##3}\VANthebibliography}

\usepackage{graphicx}	
\usepackage{amsmath}	

\title[A super-Earth planet in the WASP-84 system]{A hot super-Earth planet in the WASP-84 planetary system}

\author[G. Maciejewski et al.]{
G. Maciejewski,$^{1}$\thanks{E-mail: gmac@umk.pl}
J. Golonka,$^{1}$
W. \L{}oboda,$^{1}$
J. Ohlert,$^{2,3}$
M. Fern\'andez,$^{4}$
and F. Aceituno$^{4}$
\\
$^{1}$Institute of Astronomy, Faculty of Physics, Astronomy and Informatics,
      Nicolaus Copernicus University in Toru\'n, Grudziadzka 5, 87-100 Toru\'n, Poland\\
$^{2}$Michael Adrian Observatorium, Astronomie Stiftung Trebur, 65428 Trebur, Germany\\
$^{3}$University of Applied Sciences, Technische Hochschule Mittelhessen, 61169 Friedberg, Germany\\
$^{4}$Instituto de Astrof\'isica de Andaluc\'ia (IAA-CSIC), Glorieta de la Astronom\'ia 3, 18008 Granada, Spain\\
}

\date{Accepted for publication, 16 June 2023}

\pubyear{2023}

\begin{document}
\label{firstpage}
\pagerange{\pageref{firstpage}--\pageref{lastpage}}
\maketitle

\begin{abstract}
	Hot Jupiters have been perceived as loners devoid of planetary companions in close orbital proximity. However, recent discoveries based on space-borne precise photometry have revealed that at least some fraction of giant planets coexists with low-mass planets in compact orbital architectures. We report detecting a 1.446-day transit-like signal in the photometric time series acquired with the Transiting Exoplanet Survey Satellite (TESS) for the WASP-84 system, which is known to contain a hot Jupiter on a circular 8.5-day orbit. The planet was validated based on TESS photometry, and its signal was distilled in radial velocity measurements. The joint analysis of photometric and Doppler data resulted in a multi-planetary model of the system. With a mass of $15\, M_{\oplus}$, radius of $2\,  R_{\oplus}$, and orbital distance of 0.024 au, the new planet WASP-84~c was classified as a hot super-Earth with the equilibrium temperature of 1300 K. A growing number of companions to hot Jupiters indicates that a non-negligible part of them must have formed under a quiescent scenario such as disc migration or in-situ formation.  
\end{abstract}

\begin{keywords}
planets and satellites: detection -- planets and satellites: terrestrial planets -- planets and satellites: formation -- stars: individual: WASP-84 (BD+02 2056) -- techniques: photometric -- techniques: radial velocities
\end{keywords}



\section{Introduction}

The origins of giant planets on tight orbits, so-called hot Jupiters, remain unclear. Three main pathways are proposed to explain the formation of those planets: in situ formation, disc migration, and high eccentricity tidal migration \citep[see][for a review]{2018ARA&A..56..175D}. In the first scenario, a planet is formed via the rapid gas accretion onto a super-Earth-type rocky core. In the remaining two channels, giant planets are formed beyond the water frost line around their host stars and migrate inward. In the disc migration scenario, tidal interactions with dispersing remnants of a protoplanetary disc drive in-spiralling of a planet. High eccentricity migration assumes a planet is driven into a highly eccentric orbit by dynamical interactions with massive companions, then tightened and circularised by tidal interactions with its host star. In this violent scenario, low-mass planets are destroyed, leaving hot Jupiters in seclusion, which is supported by observations \cite[e.g.,][]{2021AJ....162..263H}.

However, there is a slowly growing sample of systems with hot Jupiters accompanied by nearby low-mass planets: WASP-47 \citep{2015ApJ...812L..18B}, Kepler-730 \citep{2018RNAAS...2..160Z,2019ApJ...870L..17C}, TOI-1130 \citep{2020ApJ...892L...7H}, WASP-132 \citep{2022AJ....164...13H}, and TOI-2000 \citep{2022arXiv220914396S}. These compact planetary architectures favour in situ formation or disc migration. \citet{2017AJ....154..106N} concluded that up to $15\%$ of hot Jupiters might have migrated in the quiescent scenario. This finding could mean that many hot Jupiters' companions remain undiscovered. Indeed, they are usually smaller than $4\, R_{\oplus}$, making them difficult to detect with current observational techniques. 

The WASP-84 system \citep{2014MNRAS.445.1114A} was identified to have an orbital configuration likely shaped by disc migration \citep{2015ApJ...800L...9A}. The host star BD+02~2056 is an active K0 dwarf of $10.6$ magnitude in the Gaia $G$ band. It is orbited by a 0.7 $M_{\rm Jup}$ hot Jupiter within 8.5 days. The orbit was found to be circular within a $2\sigma$ level and well aligned with the sky-projected orbital obliquity of $\lambda = -0.3 \pm 1.7 \degr$ and the true obliquity of $\phi = 17.3 \pm 7.7 \degr$ \citep{2015ApJ...800L...9A}. Based on single-sector photometry from the Transiting Exoplanet Survey Satellite \citep[TESS,][]{2015JATIS...1a4003R}, \citet{2022Loboda} reported on a tentative detection of a shallow 1.446-day transit-like signal as part of a project aimed at searching for low-mass planetary companions of hot Jupiters on distant and circular orbits (Maciejewski et al., in preparation). 

In this letter, we present the discovery of a transiting super-Earth planet interior to WASP-84~b.

\begin{figure*}
	\includegraphics[width=18cm]{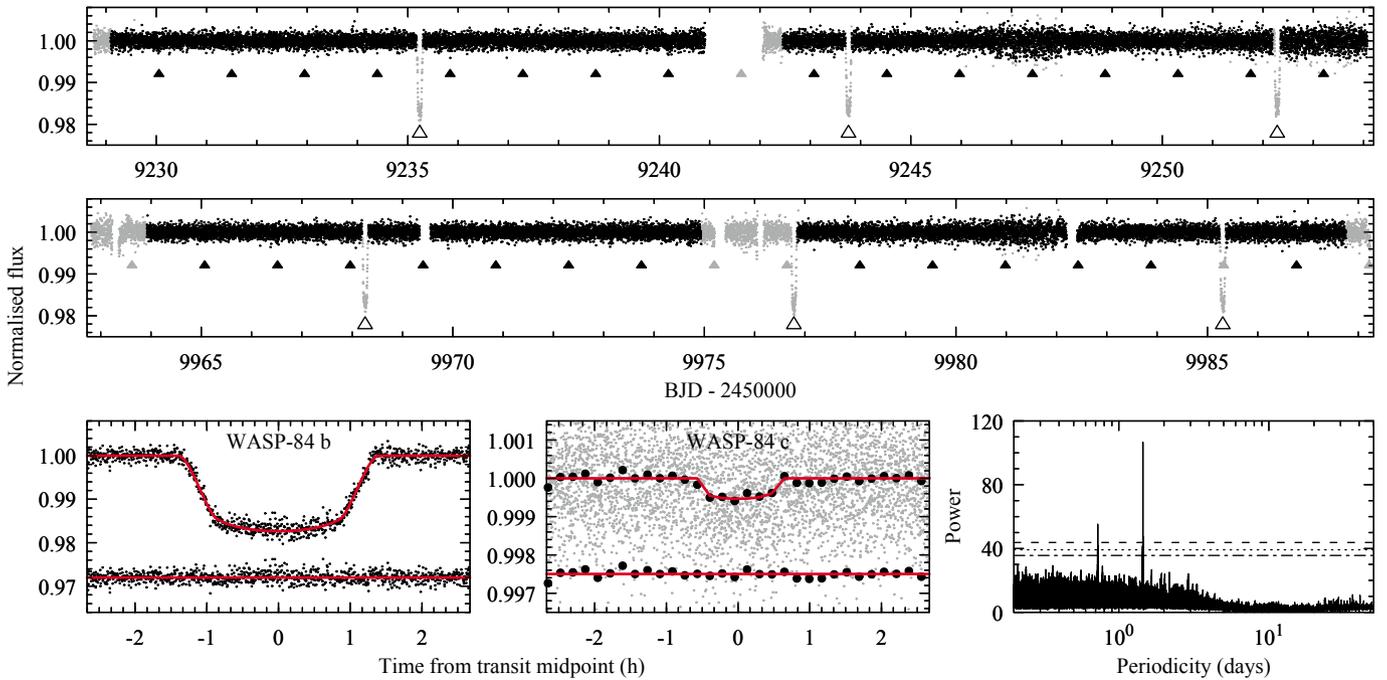}
    \caption{TESS light curve for WASP-84. \textit{Upper and middle:} Photometric time series extracted for Sector 34 and Sector 61, respectively. Grey points mark observations that were masked out in our transit search. Open and filled triangles show transits of WASP-84~b and c, respectively. The light curves before normalising are plotted in Fig.~\ref{fig:tessraw}. \textit{Bottom left:} phase-folded light curve for transits of WASP-84~b together with the best-fitting model (the red line) and the residuals plotted below. \textit{Bottom middle:} phase-folded light curve for transits of WASP-84~c. Individual measurements are shown with grey points. Black dots show data binned down to 200 points along the orbital phase. The best-fitting model is sketched with the red line. The residuals for the binned data are displayed below. \textit{Bottom right:} AoVtr periodogram calculated for Sector 34 and Sector 61 joint photometry (black points in upper panels). The horizontal lines mark the empirical FAP levels of 5\%, 1\%, and 0.1\% (from the bottom up). Periodograms calculated for each sector separately are shown in Fig.~\ref{fig:aovtr2}.}
    \label{fig:tessLC}
\end{figure*}

\section{Search for a transit-like signal} \label{Sect:Tesslc}

TESS observed WASP-84 (TIC 350293646) in Sector 34 (from UT 2021 January 14 to February 8) and 61 (from UT 2023 January 18 to February 12) within the Guest Investigator Programs G03106, G03181, G03272, G03278, G05015, G05095, and G05112. We extracted photometric time series from the $11 \times 11$ pixel ($3 \farcm 85 \times 3 \farcm 85$) wide frames acquired with the 2-minute cadence, downloaded from the MAST data archive. They were processed with the Lightkurve package \citep[ver.\ 2.0.11,][]{2018ascl.soft12013L}. The instrumental fluxes were obtained with the aperture mask, which was manually optimised to produce the lowest data-point scatter. The sky background was mapped using the standard-deviation-based thresholding method. The low-frequency trends produced by a stellar rotational variation and systematic effects were removed with the Savitzky-Golay filter with an 8-hour window and in-transit data points masked for WASP-84~b in a pre-search iteration and for both planets in the final run.

The TESS light curves (Fig.~\ref{fig:tessLC}) were cleared of measurements with quality flags other than 0. The measurements falling in transits of WASP-84~b were also removed with the 10-minute time margins. Five $\sigma$ clipping was applied to remove some sporadic outlying data points. We employed the analysis of variance algorithm optimised for the detection of box-like periodic signals \citep[AoVtr,][]{2006MNRAS.365..165S} to search for transit-like signals. We tried periods between 0.2 and 100 days with a resolution in frequency of $2 \times 10^{-5} \; {\rm day^{-1}}$. The algorithm folds the original light curve with a trial period and bins it to test a negative-pulse model with a minimum associated with a transit event. As this approach might be sensitive to the number of bins and a signal period considered, we iterated the bin number from 10 to 100 with a step of 5 to identify a periodogram with the highest peak. For the WASP-84 observations, the optimal bin number was 35, resulting in a strong peak at 1.446 days. 

A visual inspection revealed a flux drop with a depth of about 500 ppm and a duration of about 60 min. To estimate its statistical significance, we applied the bootstrap resampling method with the false alarm probabilities (FAPs) determined empirically on $10^5$ trials. We found the 1.446-day signal well above the significance level of 0.1 per cent.

We repeated the search procedure with 20-s cadence and long-cadence data (10 minutes for Sector 34 and 200 seconds for Sector 61) extracted from full-frame images downloaded with the TESSCut tool \citep{2019ascl.soft05007B}. The Box Least Squares \citep[BLS,][]{2002A&A...391..369K} algorithm was also employed as an alternative tool for detecting transit-like signals. The results of those experiments confirmed the detection of the 1.446-day signal.

\section{Validation of WASP-84~c}\label{Sect:Valid}

Using the online tool FluxCT \citep{2023RNAAS...7...18S}, we found that the flux contamination in the TESS aperture is 0.14\%. Although this contamination is negligible, the detected signal could be an astrophysical false positive resulting from a blended eclipsing binary or another planetary system. Using built-in procedures of the TRICERATOPS code \citep{2020ascl.soft02004G, 2021AJ....161...24G}, which uses a Bayesian framework for vetting and validating transiting planet candidates, we identified 4 stars within an angular distance of $63\arcsec$, i.e., 3 TESS pixels, from WASP-84. The brightest star, TIC 350290592, was 3.1 mag fainter and was separated by $58\farcm1$. The algorithm identified it as the only source bright enough to produce the observed transit-like signal. The remaining three stars were 7--9 magnitudes fainter. A trial light curve was created with a smaller aperture, in which pixels that could contain flux from nearby stars were removed. Its photometric quality was slightly worse, but it minimized flux contamination from the identified stars. The 1.446-day signal remained unchanged, so it appeared unlikely to be the astrophysical false positive originating from the resolved stars.

TRICERATOPS was fed with the trial light curve, which was phase folded and binned to 200 data points to speed up calculations. We also built a contrast curve from observations brought by \citet{2015A&A...579A.129W}. They were acquired with the Lucky Imaging camera AstraLux Norte at the Calar Alto 2.2-m telescope in the $z'$ passband. The code reported the false-positive probability (FPP) and nearby false-positive probability (NFPP) after analysing several astrophysical false positives, including eclipsing binaries and transiting planets for unresolved stars. We ran TRICERATOPS 20 times and took the means and their standard deviations as the final values. We obtained FPP of $0.0107 \pm 0.0006$ with NFPP of $0.00031 \pm 0.00001$. Candidates with ${\rm FPP} < 0.015$ and ${\rm NFPP} < 0.001$ might be classified as validated planets \citep{2021AJ....161...24G}. Thus, we conclude that the 1.446-day signal is likely to be produced by a planetary companion to WASP-84~b. We note, however, that its signal-to-noise ratio is relatively low, with a value of $15.4\pm1.6$ calculated using Eq.~17 of \citet{2021AJ....161...24G}. 

To straighten our finding, we run VESPA \citep{2012ApJ...761....6M, 2015ascl.soft03011M} to determine FPP based on stellar population simulations and several blended eclipsing binary scenarios. We provided the stellar properties of WASP-84, the effective temperature $T_{\rm eff} = 5280 \pm 80 \, {\rm K}$, surface gravity $\log g = 4.65 \pm 0.17 \, ({\rm cgs})$, and metallicity ${\rm [Fe/H]} = 0.09 \pm 0.12 \, {\rm dex}$ from \citet{2015ApJ...800L...9A}. We also provided the $BV$ and 2MASS $JHK$ magnitudes, parallax, and maximal interstellar extinction in the $V$ band. The latter parameter was extracted from the 3D dust map Bayestar19 \citep{2019ApJ...887...93G}. To determine the maximum allowed depth of a potential secondary eclipse, we searched for an eclipse-like structure of the same duration as the duration of the transit signature, located near the orbital phase of 0.5. The most profound drop we identified had 63 ppm, so we used this value as the secondary eclipse threshold. The aperture radius reflecting conservatively the aperture mask used was set to $42\arcsec$, i.e., two pixels of TESS CCDs. The contrast curve completed the VESPA's inputs.

Again, we used 20 iterations to obtain the averaged FPP that was found to be equal to $0.015 \pm 0.006$. Our result is consistent with the validation threshold of $0.01$ \citep{2016ApJ...822...86M} within the 1$\sigma$ level. Thus, we could consider WASP-84~c as being validated with high probability.

\section{System model}\label{Sect:Sysmodel}

We constructed the system model with ALLESFITTER \citep{allesfitter-code, allesfitter-paper} by joining the available photometric and Doppler time series. The TESS photometry was enhanced with two transit light curves in a $V$+$R$ filter from \citet{2014MNRAS.445.1114A}, acquired with the RISE camera on the 2-m Liverpool Telescope in January 2013, and two new transit light curves obtained by us in white light. The first was acquired with the 1.2-m Cassegrain  telescope and an SBIG STL-6303E CCD at the Michael Adrian Observatory (Trebur, Germany) on UT 2016 Mar 14. The second one was obtained with the 0.9-m Ritchey-Chr\'etien Telescope and a Roper Scientific VersArray 2048B CCD at the Sierra Nevada Observatory (OSN, Spain) on UT 2018 Dec 7. Data reduction was performed with the AstroImageJ software \citep{2017AJ....153...77C}. Fluxes were calculated after optimising the aperture radius and an ensemble of comparison stars. Trends against air mass, time, the x and y position on the chip, and seeing were removed. The final light curves were normalised to a baseline outside the transit. They are shown in Fig.~\ref{fig:bLCs} together with the literature data.

The radial velocity (RV) measurements from \citet{2014MNRAS.445.1114A} were the only ones publicly available. Those twenty RVs were acquired with the CORALIE spectrograph mounted on the Euler-Swiss 1.2-m telescope between 2011 Dec 22 and 2012 Mar 18. We noticed that these data alone were insufficient to detect the WASP-84~c signal. Thus, we added 30 RV measurements acquired with HARPS on the ESO 3.6 m telescope by \citet{2015ApJ...800L...9A}. As these measurements were publicly unavailable, we downloaded the original observations from the ESO Data Archive and reduced them with the SERVAL code \citep{2018A&A...609A..12Z}.

ALLESFITTER was fed with TESS, RISE, our new photometry, and the CORALIE and HARPS RV data. All timestamps were converted into barycentric Julian dates and barycentric dynamical time ($\rm{BJD_{TDB}}$). To speed up calculations with TESS data, the code cut 8-hour-long chunks around the transit midpoints. The stellar limb darkening (LD) was approximated with the quadratic LD law \citep{1950HarCi.454....1K} and probed in the physical $\mu$-space. The values of the linear $u_1$ and quadratic $u_2$ coefficients were interpolated from tables of \citet{2011AA...529A..75C} and varied under the Gaussian priors of a width of 0.1. The in-transit RV measurements were masked because our model did not include the orbit alignment of WASP-84~b. The free parameters of the model were the orbital periods $P$, transit epochs $T_0$, planet-to-star radii ratios $R/R_{\star}$, scaled semi-major axes $a/R_{\star}$, orbital inclinations $i$, and RV amplitudes $K$ for both planets, as well as the RV offsets and jitter terms for the CORALIE and HARPS data sets. The Mat\'ern 3/2 kernel was used to account for any variations in the photometric data. As discussed in Sect.~\ref{Sect:Discus}, no transit timing variations (TTVs) were allowed, and the orbits were assumed to be circular. To keep consistency in this multi-planet model, we provided the host's mass $M_{\star} = 0.853 \pm 0.038 \, M_{\odot}$ and radius $R_{\star} = 0.768 \pm 0.019 \, R_{\odot}$ \citep{2015ApJ...800L...9A} as the external priors on stellar density.

\begin{figure}
	\includegraphics[width=\columnwidth]{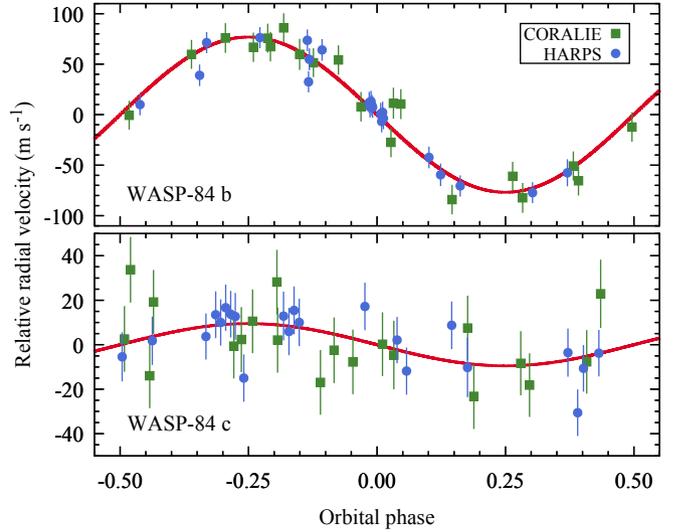}
    \caption{Phase-folded RV curves induced by WASP-84~b and WASP-84~c. The CORALIE and HARPS measurements are plotted with green squares and blue dots. The best-fitting models are drawn with red lines. The original error bars were increased by jitter determined for both datasets separately. Phase zero corresponds to transit midpoints. No correction for the stellar rotation was applied.}
    \label{fig:rvs}
\end{figure}

To explore the hyperspace of free parameters, the Nested Sampling algorithm with the default settings \citep{allesfitter-paper} was used. Prior values for these parameters can be found in Table~\ref{Table:Priors}. The parameters of the best-fitting model with their uncertainties and calculated physical properties are given in Table~\ref{Table:Results}. The transit models are plotted in Figs.~\ref{fig:tessLC} and \ref{fig:bLCs}. The RV variations split into individual components are shown in Fig.~\ref{fig:rvs}.

\begin{table}
	\centering
	\caption{Two-planetary model parameters for the WASP-84 system.}
	\label{Table:Results}
	\begin{tabular}{l c}
		\hline
Parameter (unit) & Value \\
		\hline
\multicolumn{2}{c}{\textit{Fitted parameters}} \\ 
$R_{\rm b} / R_{\star}$ & $0.12793\pm0.00058$  \\ 
$a_\mathrm{b}/R_{\star}$ & $21.78_{-0.20}^{+0.21}$ \\
$i_\mathrm{b}$ (deg) & $88.292_{-0.042}^{+0.045}$ \\ 
$T_{0,b}$ ($\mathrm{BJD_{TDB}}$) & $2457956.71194\pm0.00011$  \\ 
$P_{\rm b}$ $(\mathrm{d})$ & $8.52349648\pm0.00000060$ \\ 
$K_{\rm b}$ $(\mathrm{km\,s^{-1}})$ & $0.0768\pm0.0027$ \\ 
$R_{\rm c} / R_\star$ & $0.0233\pm0.0014$ \\ 
$a_\mathrm{c}/R_\star$ & $6.61_{-0.23}^{+0.22}$  \\ 
$i_\mathrm{c}$ (deg) & $83.20_{-0.49}^{+0.51}$  \\ 
$T_{0,c}$ ($\mathrm{BJD_{TDB}}$) & $2457952.4543_{-0.0031}^{+0.0018}$  \\ 
$P_{\rm c}$ $(\mathrm{d})$ & $1.4468849_{-0.0000016}^{+0.0000022}$  \\ 
$K_{\rm c}$ $(\mathrm{km\,s^{-1}})$ & $0.0095\pm0.0026$ \\ 
RV offset CORALIE $(\mathrm{km\,s^{-1}})$ & $-11.5784_{-0.0031}^{+0.0033}$ \\ 
RV offset HARPS $(\mathrm{km\,s^{-1}})$ & $-11.5900\pm0.0023$  \\ 
$\ln{\sigma_\mathrm{jitter, CORALIE}}$ $(\ln{ \mathrm{km\,s^{-1}} })$ & $-4.34_{-0.19}^{+0.21}$ \\ 
$\ln{\sigma_\mathrm{jitter, HARPS}}$ $(\ln{ \mathrm{km\,s^{-1}} })$ & $-4.58_{-0.15}^{+0.16}$  \\ 
\multicolumn{2}{c}{\textit{Derived parameters}} \\ 
Companion radius b, $R_\mathrm{b}$ ($\mathrm{R_{\oplus}}$) & $10.72\pm0.27$ \\ 
Companion radius b, $R_\mathrm{b}$ ($\mathrm{R_{Jup}}$) & $0.956\pm0.024$ \\ 
Semi-major axis b, $a_\mathrm{b}$ (au) & $0.0778\pm0.0021$ \\ 
Companion mass b, $M_\mathrm{b}$ ($\mathrm{M_{\oplus}}$) & $220\pm18$ \\ 
Companion mass b, $M_\mathrm{b}$ ($\mathrm{M_{Jup}}$) & $0.692\pm0.058$ \\ 
Impact parameter b, $b_\mathrm{b}$ & $0.649_{-0.011}^{+0.010}$ \\ 
Transit depth b, $\delta_\mathrm{tr,b}$ (ppth) & $17.23_{-0.12}^{+0.10}$ \\ 
Total transit duration b, $T_\mathrm{tot,b}$ (h) & $2.760\pm0.010$ \\ 
Companion density b, $\rho_\mathrm{b}$ (cgs) & $0.98_{-0.12}^{+0.13}$ \\ 
Companion surface gravity b, $g_\mathrm{b}$ (cgs) & $1901\pm87$ \\ 
Equilibrium temperature b, $T_\mathrm{eq,b}$ (K) & $732\pm12$ \\ 
Companion radius c, $R_\mathrm{c}$ ($\mathrm{R_{\oplus}}$) & $1.95\pm0.12$ \\ 
Companion radius c, $R_\mathrm{c}$ ($\mathrm{R_{jup}}$) & $0.174\pm0.011$ \\ 
Semi-major axis c, $a_\mathrm{c}$ (au) & $0.02359\pm0.00100$ \\ 
Companion mass c, $M_\mathrm{c}$ ($\mathrm{M_{\oplus}}$) & $15.2_{-4.2}^{+4.5}$ \\ 
Companion mass c, $M_\mathrm{c}$ ($\mathrm{M_{jup}}$) & $0.048_{-0.013}^{+0.014}$ \\ 
Impact parameter c, $b_\mathrm{c}$ & $0.785_{-0.047}^{+0.034}$ \\ 
Transit depth c, $\delta_\mathrm{tr,c}$ (ppth) & $0.532\pm0.057$ \\ 
Total transit duration c, $T_\mathrm{tot,c}$ (h) & $1.111_{-0.064}^{+0.075}$ \\ 
Companion density c, $\rho_\mathrm{c}$ (cgs) & $11.2_{-3.5}^{+4.5}$ \\ 
Companion surface gravity c, $g_\mathrm{c}$ (cgs) & $3800_{-1100}^{+1300}$ \\ 
Equilibrium temperature c, $T_\mathrm{eq,c}$ (K) & $1329_{-30}^{+31}$ \\ 
		\hline
	\end{tabular}
\end{table}

\section{Concluding discussion}\label{Sect:Discus}

\begin{figure}
	\includegraphics[width=\columnwidth]{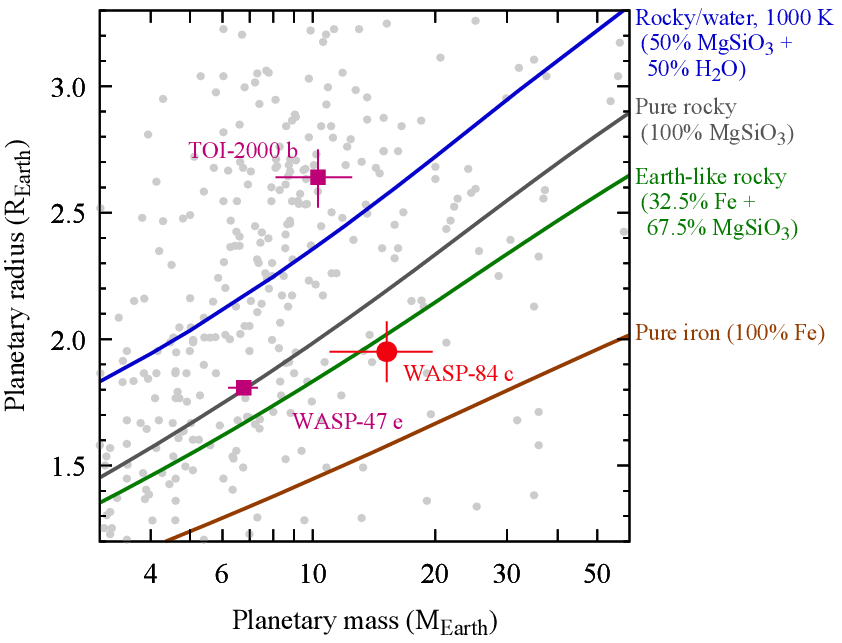}
    \caption{Planetary mass-radius diagram around WASP-84~c. The planet's location is marked with a red dot. The dark-magenta squares show two other planetary companions to the hot-Jupiters in the systems WASP-47 and TOI-2000. Their parameters were taken from \citet{2022AJ....163..197B} and \citet{2022arXiv220914396S}. The continuous lines plot the planet models of various chemical composition from \citet{2019PNAS..116.9723Z}. The background catalogue data, marked with grey small dots, were taken from the portal exoplanet.eu of The Extrasolar Planets Encyclopaedia (accessed on 2023 March 31).}
    \label{fig:diagram}
\end{figure}

With the mass of $15.2_{-4.2}^{+4.5}\,M_{\oplus}$ and radius of $1.95\pm0.12\,R_{\oplus}$, giving the mean density of $11.2_{-3.5}^{+4.5} \, {\rm g\,cm^{-3}}$, WASP-84~c can be classified as an Earth-like rocky globe. Fig.~\ref{fig:diagram} shows the location of the planet in the mass-radius diagram together with planetary models of various composition from \citet{2019PNAS..116.9723Z}. The physical properties of WASP-84~c are consistent with a model of the chemical composition of 32.3\% Fe and 67.5\% MgSiO$_{3}$.

The 1.446-day signal was not reported by the TESS Science Processing Operation Center \citep[SPOC][]{2016SPIE.9913E..3EJ} pipeline\footnote{On 2023 March 26, the 1.446-day signal was independently identified as a planet candidate in ExoFOP.}. We analysed the Presearch Data Conditioning Simple Aperture Photometry (PDC$\_$SAP) light curves produced by SPOC. In Sector 34, the photometry quality was similar to our data, and the transit signal could be detected with equal significance. This finding shows that our independent approach based on AoVtr complements other search algorithms. Compared to our data, the PDC$\_$SAP light curve in Sector 61 exhibited rms greater by a factor of 2, leaving WASP-84~c hidden in the noise.

\citet{2014MNRAS.445.1114A} observed that the host star reveals the rotational brightness modulation with an amplitude up to 3 per cent and a period $P_{\rm rot,\star} = 14.36\pm0.35$ d. Indeed, the instrumental TESS light curves display photometric variability with a periodicity of $\approx 14$ d and amplitudes of about 2--3 per cent (Fig.~\ref{fig:tessraw}). As this rotational variation was flattened by de-trending and did not affect our conclusions, we left it out of the scope of our study. We note, however, that its magnitude in the Doppler domain could be of the order of $K_{\rm c}$. \citet{2014MNRAS.445.1114A} pre-whitened the CORALIE RVs by modelling the 1-planet RV residuals with a periodic signal with a period fixed at $P_{\rm rot,\star}$. This approach was more efficient than pre-whitening using a relation between the RV residuals and bisector span. We pre-whitened the CORALIE and HARPS data sets separately, approximating the RV residuals from the 2-planet model with a fake RV keplerian signal. Its period was allowed to be fitted in a range of $14.36\pm0.35$ d. This manipulation compensated for some positive outliers in the CORALIE data (as in \citet{2014MNRAS.445.1114A}) and two negative outliers in the HARPS data. The determined RV amplitudes for both planets were more precise and consistent within $1\sigma$ with the values obtained in Sect.~\ref{Sect:Sysmodel}. As this pre-whitening did not affect our conclusions and could lead to underestimated uncertainties, we dropped it in the final iteration at the cost of accepting the higher jitter.

We conducted an additional test to determine if the 2-planet RV model has statistical support. Since random noise from stellar activity can affect RV measurements, relying on $\chi^2$-based indicators might be unreliable. To address this, we utilised the leave-one-out cross-validation routine \citep[e.g.,][]{2010arXiv1012.3754A} available in the Systemic Console Package \citep{2009PASP..121.1016M} to compare 1-planet and 2-planet models. The results favoured the 2-planet model.

Since the ratio of the orbital periods $P_{\rm b} / P_{\rm c}$ is far from a resonance, TTVs induced by mutual gravitational perturbations are expected to be weak for both planets. Our numerical experiments showed that $e_{\rm b}$ and $e_{\rm c}$ were consistent with zero within $2\sigma$. Thus, we used the circular orbits in the final model. The expected TTV amplitudes were 3 and $7\, {\rm s}$ for WASP-84~b and c, respectively. In a Newtonian trial model with the eccentricities allowed to be free, their values were found to be $e_{\rm b} \approx 0.045$ and $e_{\rm c} \approx 0.4$. The TTV amplitudes increased to $5$ and $15 \, {\rm s}$ for planet b and c, respectively. In both scenarios, such signals would be below the current detection capabilities. We employed the SWIFT package with the Regularised Mixed Variable Symplectic integrator implemented in the Systemic Console with a step of $0.01 \, P_{\rm{c}}$ to track the evolution of the orbital parameters over $10^8$ years. The system occurred to be dynamically stable for the circular and eccentric orbits.

As demonstrated by \citet{2016ApJ...825...98H}, massive planets in orbits with periods between 10 and 200 days, so-called warm Jupiters, tend to be accompanied by low-mass planets much more frequently than hot Jupiters do. This picture could manifest different formation mechanisms of both types of planetary giants. Assuming the modified tidal quality factor of $10^5$ for WASP-84~b, the circularisation time scale, calculated from Eq.~6 in \citet{2008ApJ...686L..29M}, is 12 Gyr. This is much longer than the age of the system of $2.1 \pm 1.6$ Gyr \citep{2015ApJ...800L...9A}. Thus, it is likely that the nearly circular orbit of WASP-84~b is primordial, making favouring conditions for preserving the close planetary companion. \citet{2017A&A...602A.107B} used the parameter $\alpha$, defined as the ratio of the planet's semi-major axis to its Roche limit, to identify planets with $\alpha > 5$ and circular orbits that were unlikely to undergo the high eccentricity migration. Indeed, the hot Jupiters with close companions discovered so far have $\alpha$ in a range between $4.55^{+0.24}_{-0.20}$ (WASP-47~b) and $7.44^{+0.27}_{-0.25}$ (TOI-2000~c). With $\alpha=7.18^{+0.22}_{-0.19}$, WASP-84~b occupies the outer edge of this distribution. Typical hot Jupiters ended up at $\alpha$ between 2 and 4 with a pileup at $\alpha \approx 2.5$ \citep{2017A&A...602A.107B}. As those planets seem not to be flanked by close low-mass companions, they could constitute a population distinct from $\alpha \gtrapprox 4.5$ hot Jupiters, which could be de facto a prolongation of a hot tail of the warm-Jupiter distribution \citep{2016ApJ...825...98H}.

\section*{Acknowledgements}

We thank the anonymous referee for their valuable feedback and comments on this study. GM acknowledges the financial support from the National Science Centre, Poland through grant no. 2016/23/B/ST9/00579. MF acknowledges financial support from grants PID2019-109522GB-C52/AEI/10.13039/501100011033 of the Spanish Ministry of Science and Innovation (MICINN) and PY20\_00737 from Junta de Andaluc\'{\i}a. MF and FA acknowledge financial support from the grant CEX2021-001131-S funded by MCIN/AEI/ 10.13039/501100011033. This research is partly based on data obtained at the 0.9 m telescope of the Sierra Nevada Observatory (Spain), which is operated by the Consejo Superior de Investigaciones Cient\'{\i}ficas (CSIC) through the Instituto de Astrof\'{\i}sica de Andaluc\'{\i}a. This paper includes data collected with the TESS mission, obtained from the MAST data archive at the Space Telescope Science Institute (STScI). Funding for the TESS mission is provided by the NASA Explorer Program. STScI is operated by the Association of Universities for Research in Astronomy, Inc., under NASA contract NAS 5-26555. This research made use of (1) Lightkurve, a Python package for Kepler and TESS data analysis \citep{2018ascl.soft12013L} and (2) data obtained from or tools provided by the portal exoplanet.eu of The Extrasolar Planets Encyclopaedia.

\section*{Data Availability}

The data underlying this article are available via CDS.



\bibliographystyle{mnras}
\bibliography{w84.bib} 




\appendix

\section{Additional materials}

This appendix contains additional materials that are referenced in the main body of this paper. 

\begin{figure*}
	\includegraphics[width=18cm]{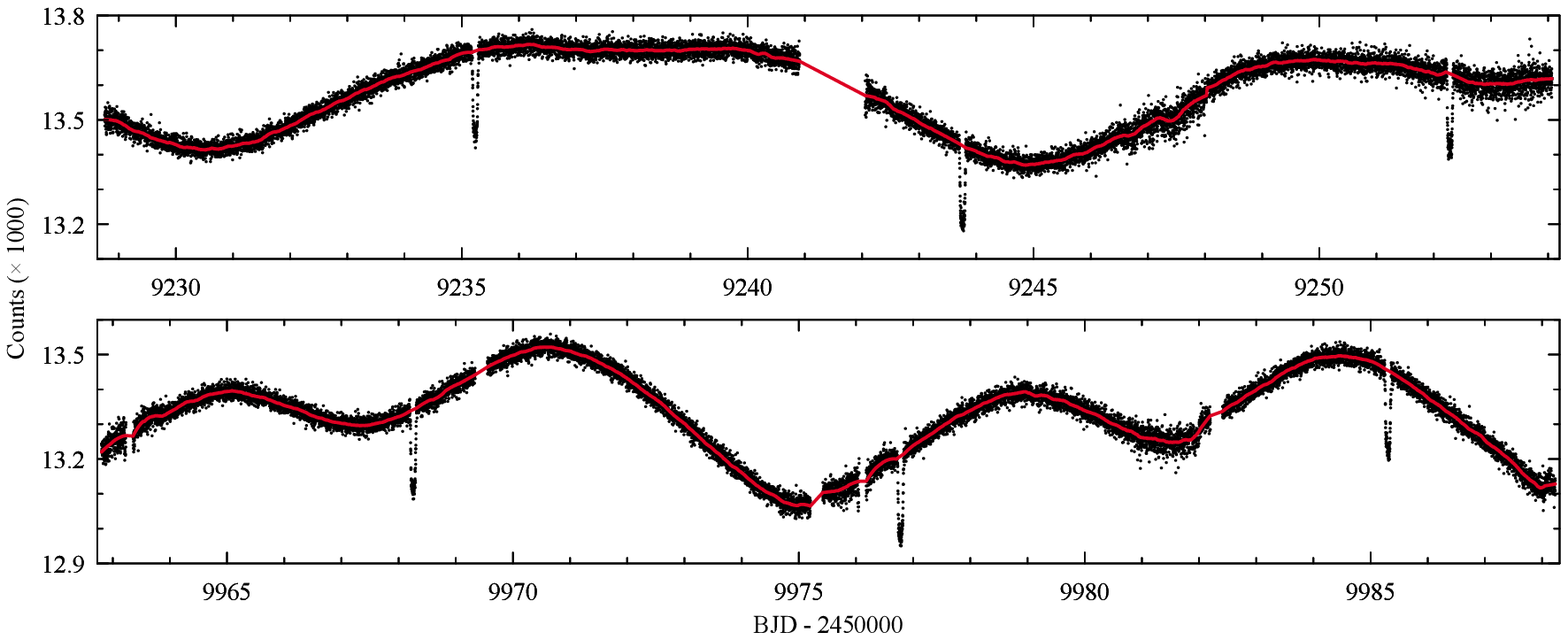}
    \caption{Raw TESS light curves for WASP-84 from simple aperture photometry. The stellar rotational variability with the period of about $14$ days is clearly visible. The shape of this modulation varies in both sectors: it is sinusoidal in Sector 34 (upper) and exhibits double minima and maxima in Sector 61 (bottom). The model used for light curve flattening is plotted with red line.}
    \label{fig:tessraw}
\end{figure*}

\begin{figure}
	\includegraphics[width=\columnwidth]{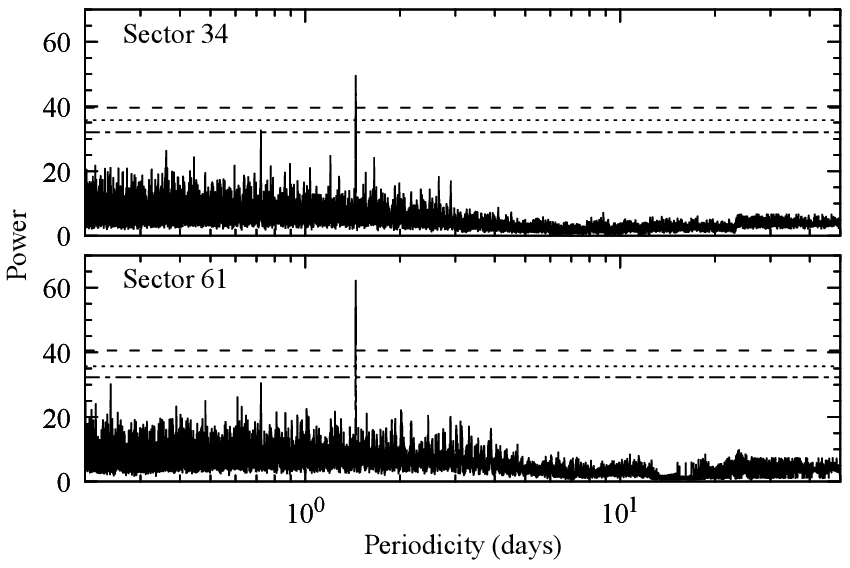}
    \caption{AoVtr periodograms calculated for data from Sector 34 and Sector 61. The horizontal lines mark the empirical FAP levels of 5\%, 1\%, and 0.1\% (from the bottom up).}
    \label{fig:aovtr2}
\end{figure}

\begin{figure}
	\includegraphics[width=\columnwidth]{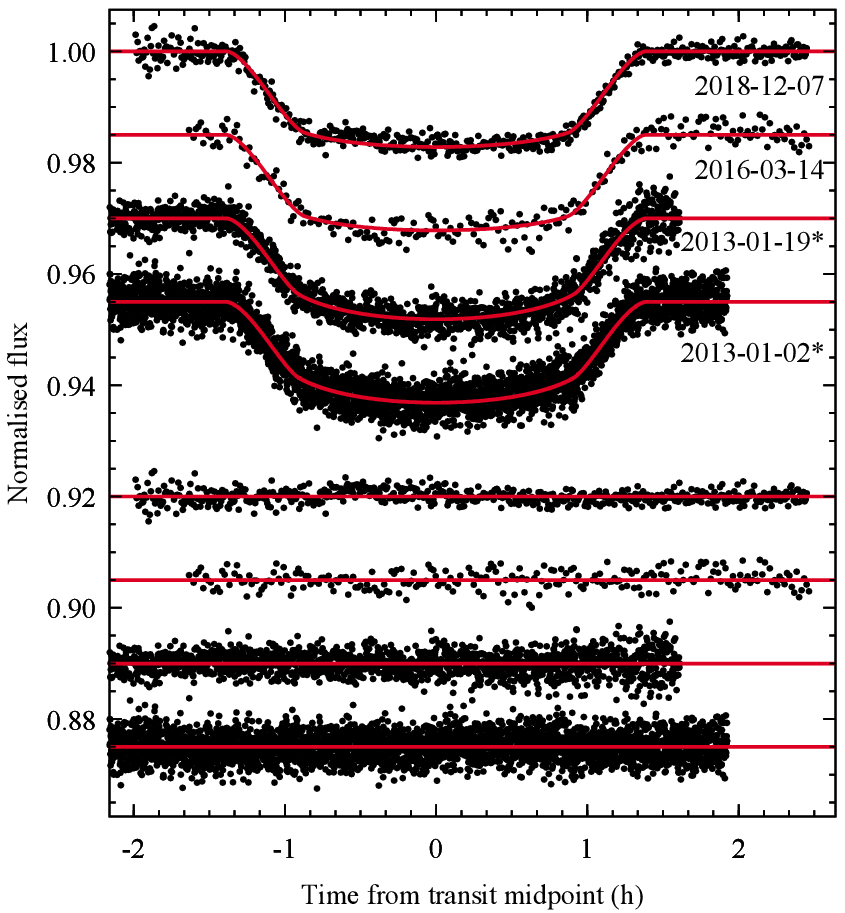}
    \caption{Ground-based light curves for transits of WASP-84~b used in this study. The two new light curves were acquired on 2016 Mar 14 and 2018 Dec 7 with the exposure times of 50 and 20 s, respectively. The photometric scatter was 1.60 and 0.77 ppth of the normalised flux per minute of observation for those nights. In the 2018 Dec 7 light curve, a flat-bottom deformation can be attributed to starspot occultation. The light curves taken from \citet{2014MNRAS.445.1114A} are also shown and indicated with an asterix. The best-fitting model is drawn with red lines, and the residuals are plotted below.}
    \label{fig:bLCs}
\end{figure}

\begin{table}
	\centering
	\caption{Parameter priors used in the model of the WASP-84 system.}
	\label{Table:Priors}
	\begin{tabular}{l c}
		\hline
Parameter (unit) & Prior \\
		\hline
$R_{\rm b} / R_{\star}$ & $\mathcal{U}(0.120, 0.135)$  \\ 
$a_\mathrm{b}/R_{\star}$ & $\mathcal{U}(18, 26)$ \\
$i_\mathrm{b}$ (deg) & $\mathcal{U}(87, 89)$ \\ 
$T_{0,b}$ ($\mathrm{BJD_{TDB}}-2450000$) & $\mathcal{U}(7956.706, 7956.718)$  \\ 
$P_{\rm b}$ $(\mathrm{d})$ & $\mathcal{U}(8.52345, 8.52355)$ \\ 
$K_{\rm b}$ $(\mathrm{km\,s^{-1}})$ & $\mathcal{U}(0.045, 0.115)$ \\ 
$R_{\rm c} / R_\star$ & $\mathcal{U}(0.001, 0.052)$ \\ 
$a_\mathrm{c}/R_\star$ & $\mathcal{U}(2.5, 12.0)$  \\ 
$i_\mathrm{c}$ (deg) & $\mathcal{U}(60, 90)$  \\ 
$T_{0,c}$ ($\mathrm{BJD_{TDB}}-2450000$) & $\mathcal{U}(7952.4, 7952.5)$  \\ 
$P_{\rm c}$ $(\mathrm{d})$ & $\mathcal{U}(1.4460, 1.4476)$  \\ 
$K_{\rm c}$ $(\mathrm{km\,s^{-1}})$ & $\mathcal{U}(0.001, 0.025)$ \\ 
RV offset CORALIE $(\mathrm{km\,s^{-1}})$ & $\mathcal{U}(-11.66, -11.50)$ \\ 
RV offset HARPS $(\mathrm{km\,s^{-1}})$ & $\mathcal{U}(-11.66, -11.50)$  \\ 
$\ln{\sigma_\mathrm{jitter, CORALIE}}$ $(\ln{ \mathrm{km\,s^{-1}} })$ & $\mathcal{U}(-15, 0)$ \\ 
$\ln{\sigma_\mathrm{jitter, HARPS}}$ $(\ln{ \mathrm{km\,s^{-1}} })$ & $\mathcal{U}(-15, 0)$  \\ 
$u_{\rm 1,TESS}$ & $\mathcal{N}(0.377, 0.100)$ \\
$u_{\rm 2,TESS}$ & $\mathcal{N}(0.235, 0.100)$ \\
$u_{\rm 1,clear}$ & $\mathcal{N}(0.464, 0.100)$ \\
$u_{\rm 2,clear}$ & $\mathcal{N}(0.216, 0.100)$ \\
$u_{\rm 1,RISE}$ & $\mathcal{N}(0.515, 0.100)$ \\
$u_{\rm 2,RISE}$ & $\mathcal{N}(0.205, 0.100)$ \\
$M_{\star}$ $(M_{\odot})$ & $\mathcal{N}(0.853, 0.058)$ \\
$R_{\star}$ $(R_{\odot})$ & $\mathcal{N}(0.768, 0.019)$ \\
		\hline
\multicolumn{2}{l}{Note. $\mathcal{U}(a,\, b)$ is the uniform distribution over the interval $[a,\,b]$.} \\ \multicolumn{2}{l}{$\mathcal{N}(\mu,\,\sigma)$ is the normal distribution with the mean $\mu$ and standard deviation $\sigma$.} \\ 
	\end{tabular}
\end{table}



\bsp	
\label{lastpage}
\end{document}